\documentclass{aa}
\usepackage{txfonts}
\usepackage{epsfig}
\usepackage{times}
\usepackage{natbib}
\bibpunct{(}{)}{;}{a}{}{,}
\defcitealias{Bauer01}{B01}
\begin{document}

\title{Flux and spectral variations in the Circinus Galaxy}
\author{S. Bianchi\inst{1}, G. Matt\inst{1}, F. Fiore\inst{2}, A.C.
Fabian\inst{3}, K. Iwasawa\inst{3}, F. Nicastro\inst{4} }

\offprints{Stefano Bianchi, \email{bianchi@fis.uniroma3.it}}

\institute{Dipartimento di Fisica, Universit\`a degli Studi Roma Tre, Via
della Vasca Navale 84, I-00146 Roma, Italy
\and Osservatorio Astronomico di Roma, Via dell'Osservatorio, I-00044
Monteporzio Catone Italy
\and Institute of Astronomy, Madingley Road, Cambridge CB3 0HA, UK
\and Harvard-Smithsonian Center for Astrophysics, 60 Garden Street, Cambridge
MA 02138, USA }

\date{Received / Accepted}

\authorrunning{S. Bianchi et al.}

\abstract{We report on a dramatic flux ($\sim50\%$ increase in the LECS and
MECS band) and spectral variation between two BeppoSAX observations of the
Circinus Galaxy performed almost three years apart. Through the analysis of all
$Chandra$ observations available in the archive, including a new DDT observation
on May 2001, we show that a high flux state of an extremely variable Ultra
Luminous X-ray source \citep[CG~X-1: ][]{Bauer01}, which is within the adopted
BeppoSAX source extraction region of $2\arcmin$, is the most likely explanation 
for most of the observed variation. However, the presence of a high flux 6.7 keV
line and the spectral variation of the PDS in the new BeppoSAX data could be
partly due to intrinsic variation of the nucleus. Comparing the longest
Chandra observation and the BeppoSAX one, we find that the long-term flux
variability of CG~X-1 is not accompanied by a significant spectral variability.
We also re-analysed the $Chandra$ HEG nuclear spectra and report on the presence
of a Compton shoulder with a flux of about $20\%$ the line core, in agreement
with theoretical expectations for Compton-thick matter.


}

\maketitle

\section{Introduction}

The Circinus Galaxy hosts one of the closest (3.8 Mpc) and X-ray brightest
($F_{\rm 2-10\,keV} \sim 1.5\times10^{-11}$ erg cm$^{-2}$ s$^{-1}$) Seyfert~2.
The detailed HST images clearly reveal a compact ($<2$ pc) active nucleus  seen
through high obscuration, surrounded by extended and complex structures
\citep{wil00}.

The first X-ray observation was performed during the $ROSAT$ All Sky Survey
\citep{brink94}. A reflection dominated spectrum was revealed by $ASCA$,
together with a prominent neutral iron K$\alpha$ line and a number of other
lines from lighter elements \citep{Matt96}. The BeppoSAX observation added a
precious piece of information, detecting the direct nuclear emission above
$\sim10$ keV, absorbed at lower energies by a Compton--thick ($4\times10^{24}$
cm$^{-2}$) material, usually identified with the torus envisaged in unification
models \citep{Matt99, Guainazzi99}.

The line spectrum above $\sim2$ keV clearly originates from low ionized
matter \citep{Matt96,net98,sako00}. It was shown that it is fully compatible
with reflection from the inner surface of a mildly ionized torus, the same
matter likely responsible for the absorption of the nuclear radiation: this
interpretation leads to an estimate of the inner radius of the torus of
$\sim0.2$ pc \citep{bmi01}. The spectrum below $\sim2$ keV is probably
contaminated by an extended emission and/or off-nuclear sources within $ASCA$
and BeppoSAX extraction regions.

This scenario was basically confirmed by $Chandra$. Two different regions are
clearly observed: one compact and spatially unresolved ($<$15 pc) which
corresponds to the nucleus, where the reflection spectrum and the iron line is
produced; the other, extended over about 50 pc, where most of the soft emission
is produced \citep{Sambruna01a,Sambruna01b}. Furthermore, a number of
off-nuclear sources were detected, mostly concentrated within $2\arcmin$ of the
nucleus \citep{sw01,Bauer01,Sambruna01a}: at least one of them is likely to have
contaminated $ASCA$ and SAX observations during its high flux states (see
below).

In this paper we report on a dramatic flux and spectral variation detected in a
second BeppoSAX observation of the Circinus Galaxy performed on January 2001,
almost three years later than the first one. In order to discriminate between a
nuclear variation and the contamination of an off--nuclear source, we will make
extensive use of all $Chandra$ observations available in the archive, together
with unpublished Director's Discretionary Time (DDT) data taken on May 2001. We
adopt the names after \citet{Bauer01} (from now on B01) for the two sources
discussed in this paper, CG~X-1 ($\alpha_{\rm 2000}=14^h13^m12.^s3$,
$\delta_{\rm 2000}=-65\degr20\arcmin13\arcsec$) and CG~X-2 ($\alpha_{\rm
2000}=14^h13^m10.^s0$, $\delta_{\rm 2000}=-65\degr20\arcmin44\arcsec$).

\section{Observations and data reduction}

\subsection{ROSAT}

The Circinus Galaxy was observed once by $ROSAT$/PSPC in 1990 within the
All-Sky Survey \citep[e.g. ][]{vbd96}. Later on, it was observed five times by
$ROSAT$/HRI (see Table \ref{rosat}). Event files were downloaded from the HEASARC
archive and analysed with HEAsoft 5.1.

\subsection{BeppoSAX}

The Circinus Galaxy was observed for the first time by BeppoSAX
\citep{Boella97} on 1998 March 24 and then again on 2001 January 7. Data from
the first observation are discussed by \citet{Guainazzi99} and
\citet{Matt99}: we defer the reader to these papers for details.

Data from three instruments (LECS, MECS and PDS) aboard BeppoSAX will be
discussed here. The MECS works in the 1-10 keV energy band, with an energy
resolution of $\simeq8\%$ and an angular resolution of $\simeq0.7$ arcmin (FWHM)
at 6 keV. The LECS covers a larger energy band, extending down to 0.1 keV, with
characteristics similar to the MECS in the overlapping band. The PDS is a
passively collimated detector with a field of view of approximately 1.5x1.5
degrees, working in the range 13-200 keV. The effective exposure times of the
2001 observation were $2.7\times10^{4}$ s (LECS), $5.2\times10^{4}$ s (MECS)
and  $3.8\times10^{4}$ s (PDS).

As discussed in detail in \citet{Matt99}, there is a second bright source within
$5\arcmin$ from the Circinus Galaxy in the LECS and MECS images. Following
\citet{Matt99}, we avoided contamination from this source extracting
data for the LECS and the MECS in a conservative radius of $2\arcmin$ centered
on Circinus. A smaller radius implies a large fraction of lost photons: for
$1.5\arcmin$, a $20\%$ loss of the total counts is measured in the MECS. A
background was created using blank sky spectra from the same region for the
detector field of view and then subtracted. PDS data have been reduced following
the standard procedures described in \citet{Matt97}. The resulting net count
rates are shown in Table \ref{saxcount}. Data were reduced with HEAsoft 5.1 and
spectra analysed with \textsc{Xspec} 11.1.0.

\begin{table*}

\caption{\label{saxcount}Net count rates and fluxes for the two BeppoSAX 
observations.}

\begin{center}
\begin{tabular}{lllll}
\hline  & \textbf{LECS $(\times10^{-2})$} & \textbf{MECS $(\times10^{-1})$} &
\textbf{PDS} & \textbf{Flux 2-10 keV (cgs)}\\
\hline 1998-03-13 & $4.41\pm0.08$ & $1.050\pm0.009$ & $2.01\pm0.04$ &
$1.4\times10^{-11}$\\
2001-01-07 & $6.76\pm0.16$ & $1.511\pm0.017$ &
$1.93\pm0.04$ & $2.1\times10^{-11}$\\
\hline
\end{tabular}
\end{center}

\end{table*}

\subsection{Chandra}

Following the discovery of a large flux and spectral variation in the second
SAX observation (see Sect. \ref{analysis}), a $Chandra$ DDT observation was
performed on 2001 May 2. The arcsec spatial resolution of $Chandra$ allows us to
resolve several different sources within the $2\arcmin$ BeppoSAX adopted source
extraction region. Circinus had been already observed several times by $Chandra$
and a systematic study of the variability for all the sources present in the
field over a period of 1.5 years is possible (Table \ref{circlog}).

All observations were performed with the Advanced CCD Imaging Spectrometer
(ACIS-S: Garmire et al., in preparation). Only two (they are actually almost
contiguous segments) used the High-Energy Transmission Grating Spectrometer
(HETGS: Canizares et al., in preparation). Different configurations of subarray
windows, resulting in different frame times, were selected, depending on the
aims of the observations. A lower frame time reduces significantly the pileup of
the nucleus, at the expense of a less efficient use of the exposure time and
consequent lower statistics for all sources. Data were reduced with the Chandra
Interactive Analysis of Observations software (CIAO 2.2.1), using the Chandra
Calibration Database (CALDB 2.10). Spectra were analysed with \textsc{Xspec}
11.1.0, while grating data with $Sherpa$ 2.2.1. In the following, errors are at
the 90\% level of confidence for one interesting parameter ($\Delta \chi^2
=2.71$).

\begin{table*}

\caption{\label{circlog}The log of the $Chandra$ observations of the Circinus
Galaxy, along with 0.1-10 keV count rates (in units of $10^{-3}$ count
s$^{-1}$ for nucleus and CG~X-1. Flux estimates are not included because the 
nucleus is largely affected by pileup (see text for details). }

\begin{center}

\begin{tabular}{lllllll}
\hline
\textbf{Date} & \textbf{Instr}. & \textbf{Frame Time (s)} & \textbf{Exp (s)}&
\textbf{Reference}&\textbf{Nucleus}&\textbf{CG~X-1}  \\
\hline
2000-01-16 & ACIS-S & 3.2 & 964 & \citealt{sw01}& $187\pm14$ & $35\pm5$ \\
2000-03-14 & ACIS-S & 0.4 & 4974 & \citealt{sw01}& $329\pm9$ & $339\pm8$ \\
2000-03-14 & ACIS-S & 3.2 & 23076 & \citealt{sw01}& $190\pm3$ & $133\pm2$ \\
2000-06-15 & ACIS-S HETG & 2.1 & 7122 & \citealt{Bauer01}& $85\pm1$ & $16\pm1$ \\
2000-06-16 & ACIS-S HETG & 2.1 & 60222 &
\citealt{Bauer01,Sambruna01a,Sambruna01b}& $89\pm4$ & $12\pm1$ \\
2001-05-02 & ACIS-S & 3.2 & 4401 & This paper & $192\pm7$ & $67\pm4$\\
\hline
\end{tabular}

\end{center}

\end{table*}

\section{\label{analysis}Data analysis}

\subsection{Flux}

\subsubsection{\label{saxflux}BeppoSAX}

The net count rates for the new observation are significantly higher
than those of the 1998 one for the LECS and the MECS by a factor of
$\simeq50\%$. In contrast, the count rate for the PDS is slightly lower in the
new observation, even if marginally consistent with that measured in the old one
(Table \ref{saxcount}).

To avoid possible systematic errors in the effective count rates of
the new data, we performed a series of tests. First of all, we checked that the
backgrounds for the two instruments were the same in the two observations. Then
we looked at the lightcurves of the two sets of data in order to understand if
the calculated mean count rates could be contaminated by short periods of
anomalous high or low flux. Both tests indicate that the flux variation
between the two observations is a real effect.

Another possible source of contamination could be the second source visible
in the MECS and LECS images: a rise in its flux from the previous
observation could affect the Circinus flux even within the conservative
$2\arcmin$ region adopted. We then extracted the spectrum in a radius of
$1\arcmin$ centered on the second source both for the LECS and the MECS. The
net count rates are indeed higher for both instruments in the 2001 observation,
but the change is not statistically significant ($1.28\pm0.06\times10^{-2}$
against $1.21\pm0.03\times10^{-2}$ for the MECS and $6.7\pm1.7\times10^{-3}$
against $5.4\pm0.6\times10^{-3}$ for the LECS). It seems more reasonable that
it is indeed the higher flux of Circinus that contaminates the second source,
and not the contrary.

\begin{figure}
\centerline{\epsfig{figure=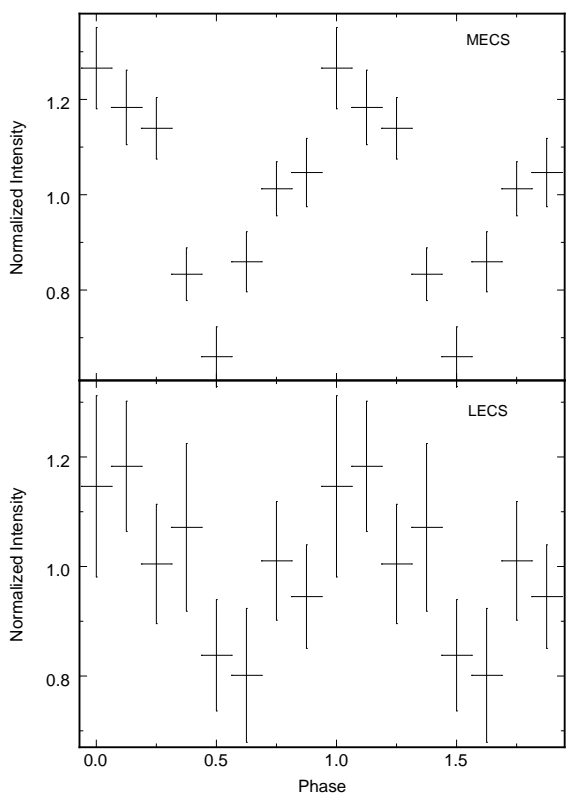,width=7.5cm}}
\caption[]{\label{folded}MECS (top) and LECS (bottom) soft band ($<3$
keV) folded lightcurve over a period of 27 ks (7.5 hr) for the January 2001
observation. The intensities were normalized on the mean value.}
\end{figure}

The lightcurve of the new data shows a larger variability than the old one. In
particular, the MECS data in the soft band (below 3 keV) seems to vary
periodically, as expected if the observed excess has to be ascribed
to a high-flux state of an off-nuclear source, CG~X-1, which is strongly
periodic (\citetalias{Bauer01}; see Sect. \ref{natureJ}).
To test this possibility, we folded the MECS soft band lightcurve with the
period of CG~X-1 (27 ks): the result, shown in Fig. \ref{folded}, clearly
indicates the presence of such a periodicity, with an amplitude between maximum
and minimum which is $\sim50\%$ of the mean value. A similar behaviour
is found in the LECS data, when folded with the same period: the
amplitude is of the same order of that observed in the MECS data (Fig.
\ref{folded}). Evidence of the same periodicity is also present in the old
data, but the variation amplitude is less, indicating a contamination from
CG~X-1 by only a factor of $\sim10\%$, in agreement with the low flux state in
which the source was generally observed by Chandra (see Sect.
\ref{chandraflux}).

\subsubsection{\label{chandraflux}Chandra}

We analysed all the bright sources within $2\arcmin$ of the nucleus
\citepalias[see Fig. 1 in ][]{Bauer01} in all $Chandra$ observations (see Table 
\ref{circlog}). To establish if the nucleus has remained constant, it must be
noted that a direct comparison between observations with different frame time is
not possible, because of the high pileup the source suffers with high frame
times. The count rates for the nucleus in the three observations with frame time
3.2 s (which span a period of 1.5 years) are remarkably constant with one
another. Moreover, the two contiguous observations with frame time 2.1 s are
also consistent to each other. The only observation free of pileup is that with
frame time 0.4 s (5\% according to
WebPIMMS\footnote{http://heasarc.gsfc.nasa.gov/Tools/w3pimms.html}).

Unfortunately, it is not possible to compare the 2.1 s frame time observations
with the others, because WebPIMMS does not support HETGS count rates with frame
time other than 3.2 s. On the other hand, if we use the count rate for the 0.4 s
observations to calculate the expected count rate with 3.2 s, we get a
value somewhat lower than the observed one, but the predicted pileup is higher
than 40\%, where WebPIMMS results are probably no longer reliable.
Therefore, even if, when comparison may be done, no evidence for variations is
found, it is not possible to establish that the nucleus remained constant in all
the observations, due to the above-mentioned problems.

All the analysed sources are always much fainter than the nucleus but CG~X-1,
which, on March 2000, reached the count rate of the AGN (Table \ref{circlog}). In
all the other observations, instead, its count rate is much lower than that of
the nucleus. The flux observed during its peak in March ($F_{\rm 0.1-10\,keV}=6
\times 10 ^{-12}$ erg cm$^{-2}$ s$^{-1}$) is fully consistent with the LECS and
MECS excess in the second SAX observation.

\subsection{\label{spectral}Spectral analysis}

\subsubsection{\label{saxspectral}BeppoSAX}

\begin{figure}
\centerline{\epsfig{figure=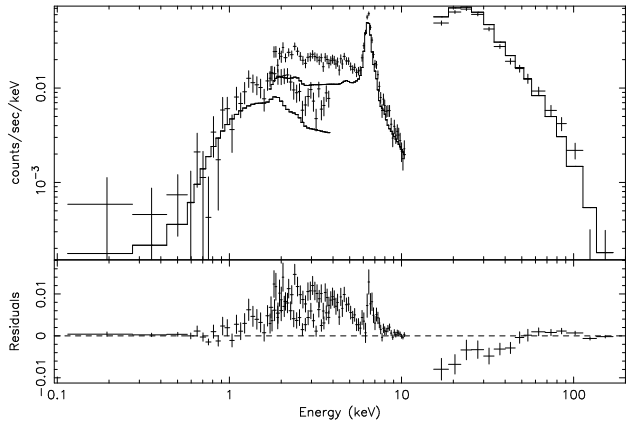,width=5.5cm,angle=-90}}
\caption[]{\label{new2old}Data from the new BeppoSAX observation superimposed
on the best fit for the old one.}
\end{figure}

Figure \ref{new2old} shows the data from the new observation superimposed on the
best fit for the old one \citep{Matt99,Guainazzi99}. A huge excess both in the
LECS and MECS is evident, together with large residuals around the iron line
energy and a decrease in the PDS. We subtracted each other the two BeppoSAX
spectra (PDS excluded) and performed a fit on the resulting spectrum. An
absorbed powerlaw ($\Gamma=2.3 ^{+0.5} _{-0.4}$) is a good model (see Table
\ref{Jfit}), together with a huge (EW=1.9 keV) iron line at
$6.67^{+0.10}_{-0.14}$ keV. The required column density ($N_\mathrm{H} =
2.0^{+1.7} _{-0.8} \times 10^{22}$ cm$^{-2}$) is higher than the Galactic one
\citep[$3.0\times 10^{21}$ cm$^{-2}$: ][]{Bauer01}. The 0.1--10 keV flux for
this model is $5.1 \times 10^{-12}$ erg cm$^{-2}$ s$^{-1}$ and the line flux is
$7.8 ^{+3.2}_{-3.0}\times 10^{-5}$ ph cm$^{-2}$ s$^{-1}$. Alternatively, the
spectrum can be fitted equally well by a multicolour disk blackbody emission
(see Table \ref{Jfit}), with an inner-disk temperature of
$T_\mathrm{in}=1.7^{+0.4}_{-0.3}$ keV and $N_\mathrm{H}=1.0^{+0.9}_{-0.6}\times 
10^{22}$ cm$^{-2}$. A 6.7 keV line with the same flux as for the previous model
is still required.

\subsubsection{\label{chandraspec}The Chandra spectrum of CG~X-1}

We then analysed the spectrum of CG~X-1 in its high flux observation, in
March 2000 (Table \ref{circlog}), in order to see if the spectral shape of the
excess in the LECS and MECS data is compatible with the spectrum of CG~X-1 as
observed by $Chandra$. Because of the selected frame time of 0.4 s, this
observation is almost free of pileup for CG~X-1 (5\% according to
WebPIMMS). We get a good fit ($\chi^2=77/92$ d.o.f.) with an absorbed powerlaw,
$N_\mathrm{H} = 1.1^{+0.1} _{-0.2} \times 10^{22}$ cm$^{-2}$ and $\Gamma = 1.4 ^{+0.2}
_{-0.1}$ \citep[in good agreement with][]{sw01}. No iron line is required by the
data: an upper limit to the flux of a 6.67 keV gaussian line is $4.1 \times
10^{-6}$ ph cm$^{-2}$ s$^{-1}$, which is not compatible with that measured by
SAX.

A multicolour disk blackbody emission fits the spectrum equally well
($\chi^2=81/92$ d.o.f.), with an inner-disk temperature of
$T_\mathrm{in}=2.4^{+0.4}_{-0.3}$ keV and $N_\mathrm{H} = 8.1^{+0.9} _{-0.8} 
\times 10^{21}$ cm$^{-2}$. This kind of spectrum represents emission from an
optically thick accretion disk around a black hole, generally a good model for
Ultra-Luminous X-ray sources \citep{Mak00}.

In order to see if the flux variation of this source is associated
with spectral variations, we extracted the spectra of CG~X-1 for all $Chandra$
observations. We decided to exclude from our analysis the second observation of
March 2000, because the combination of a high flux state of the source and a
frame time of 3.2 strongly affects its spectrum with pileup, and the shorter one
of June 2000, because the source was in the same flux state as the contiguous
observation, which was longer by a factor 10 (see Table \ref{circlog}). In each
case, we tried the two models described above, a simple powerlaw and a disk
black body. The results are summarized in Table \ref{Jfit}.

\begin{table*}

\caption{\label{Jfit}Best fit parameters for CG~X-1 for the two models
adopted in the selected $Chandra$ observations and the BeppoSAX spectrum (see
text for details). The reported fluxes are relative to the powerlaw model.}

\begin{center}

\begin{tabular}{llllllll}
\hline \textbf{Parameter} & \textbf{BeppoSAX} &\textbf{JAN00} & \textbf{MAR00a}
& \textbf{JUN00b} & \textbf{MAY01} \\
\hline $N_\mathrm{H}$ (cm$^{-2}$) & $2.0^{+1.7} _{-0.8}\times10^{22}$
&$5.6^{+0.5}_{-0}\times10^{21}$ & $1.1^{+0.1}_{-0.2}\times10^{22}$ &
$1.3^{+0.3}_{-0.2}\times10^{22}$ & $8.5^{+4.4}_{-2.6}\times10^{21}$ \\

$\Gamma$ & $2.3 ^{+0.5} _{-0.4}$ & $1.5^{+0.7}_{-0.5}$ &
$1.4^{+0.2}_{-0.1}$ & $2.5\pm0.2$ & $1.2^{+0.4}_{-0.2}$ \\

$\chi^2$ & 15 (35 dof) & 9 (7 dof) & 77 (92 dof) & 67 (56 dof) & 27 (17 dof) \\

\hline $N_\mathrm{H}$ (cm$^{-2}$) & $1.0^{+0.9}_{-0.6}\times10^{22} $
&$5.6^{+0.5}_{-0}\times10^{21}$ & $8.1^{+0.9}_{-0.8}\times10^{21}$ &
$6.8^{+1.5}_{-1.2}\times10^{21}$ & $6.9^{+2.9}_{-1.6}\times10^{21}$ \\

$T_\mathrm{in}$ (keV) & $1.7^{+0.4}_{-0.3}$ &$2.0^{+2.2}_{-0.8}$ & 
$2.4^{+0.4}_{-0.3}$ & $1.35^{+0.13}_{-0.12}$ & $2.9^{+1.3}_{-0.8}$ \\

$\chi^2$ & 16 (35 dof) &9 (7 dof) & 81 (92 dof) & 60 (56 dof) & 21 (17) \\

\hline Flux 0.1--10 keV (cgs) & $5.1\times10^{-12}$ & $5.8\times10^{-13}$ &
$6\times10^{-12}$ & $9\times10^{-13}$ & $1.2\times10^{-12}$ \\

\hline

\end{tabular}

\end{center}

\end{table*}

There is indeed evidence of spectral variations, between hard and soft states,
in terms either of different photon indexes or disk temperatures.
\citetalias{Bauer01} found that in the June observation there is a possible
hardening of the spectrum during the rising phase of the period, suggesting that
it could be the effect of an absorption at soft energies associated with that
phase. Because of the short exposure times of the other observations, it is not
possible to confirm this behaviour and ascribe the spectral variability to it.
The only observation which is long enough is the second one performed in March
but, because of the high flux state and the selected frame time, it is piled
up so heavily that it lacks significant spectral informations. However, it
should be noted that the only observation in which the spectrum appears soft is
the longest one, where we analyse the mean spectral shape of all phases of the
period. Moreover, this spectrum is fully compatible with that of the SAX
residuals, as expected since the BeppoSAX observation (an exposure time of 51 ks
distributed over 60 hours) also includes a number of periods of CG~X-1.

\subsubsection{The Compton shoulder}

We re-analysed the HEG spectrum of the nucleus to look for a Compton shoulder in
the iron K$\alpha$ line, which is not reported in \citet{Sambruna01b}, possibly
because the response matrix included with the version of CIAO available at that
time was less accurate than the new one. However, this feature is predicted on
theoretical grounds as a product of Compton scattering of the fluorescent line
photons from optically thick matter \citep[e.g.][]{gf91,matt91,rey94,Matt02}.
\citet{ifm97} observed the Compton shoulder in the $ASCA$ spectrum of NGC~1068.
Later on, \citet{kaspi02} detected this feature in the $Chandra$ HETG spectrum
of NGC~3783. We expected it to be visible in the 60ks $Chandra$ grating
observation.

\begin{figure}
\centerline{\epsfig{figure=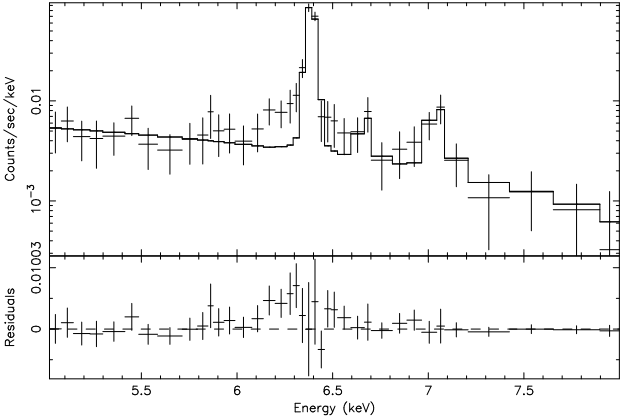,width=5.5cm,angle=-90}}
\caption[]{\label{shoulder}The Compton shoulder in the $Chandra$ HEG
spectrum showed as residuals from the best fit including three gaussian lines.}
\end{figure}

Indeed, residuals in excess of the red wing of the 6.4 keV iron line are clearly
visible in the data, if we use a powerlaw as the continuum between 5 and 8 keV
and put three gaussian lines at 6.40, 6.67 and 7.04 keV, as required by the
data (Fig. \ref{shoulder}).  We tried to fit this feature with the
\textsc{box1D} function in $Sherpa$, which has a constant value in a
given energy range and is identically null outside. This is a
reasonable approximation of the theoretical profile \citep[e.g. Fig. 2 in
][]{Matt02}. The best fit boundaries for the box are 6.15 keV and 6.40 keV,
while the flux is $6.2\pm2.0\times10^{-5}$ ph cm$^{-2}$ s$^{-1}$, that is
$25\pm8\%$ that of the iron K$\alpha$ line. If we keep the box boundaries fixed
at the values expected from theoretical models (see below), the flux of the
Compton shoulder is $5.1\pm1.8\times10^{-5}$ ph cm$^{-2}$ s$^{-1}$, that is
$20\pm7\%$ that of the iron K$\alpha$ line core.

These values are fully consistent with the theoretical expectations. If it
originated by reflection from a Compton-thick material ($N_\mathrm{H}>10^{24}$
cm$^{-2}$), the first order Compton shoulder is expected to extend from 6.2436
keV to the core of the iron line, with a centroid at 6.31-6.32 keV and a flux of
$15-20\%$ that of the core \citep{Matt02}.

As a further test, we also fitted the Compton shoulder with \textsc{Xspec},
adopting a simple gaussian line. We obtain a good correction of the residuals,
with a centroid at $E_0=6.27$ keV and a FWHM
$\sigma=0.07$ keV. The flux of the Compton shoulder modelled in
this way ($5.8^{+2.3}_{-2.6}\times10^{-5}$ ph cm$^{-2}$ s$^{-1}$) is fully
compatible with that measured with $Sherpa$.

\section{Discussion}

\subsection{\label{natureJ}The nature of CG~X-1}

Off-nuclear Ultra-Luminous X-ray sources (ULXs) have been observed in several
nearby galaxies by $Einstein$, $ROSAT$, $ASCA$ and now by $Chandra$
\citep[e.g.][]{cm99,miz01,lapa01,fab01}. Their luminosities are in the range
$10^{39}-10^{40}$ erg s$^{-1}$, largely super-Eddington for stellar mass
objects. Even if some of them can be explained as the product of young supernova
remnants \citepalias[e.g.][ and references therein]{Bauer01}, the basic idea is
that such luminosities are achieved through accretion onto black-holes with
masses of 50-500 M$_{\sun}$, i.e. intermediate between the masses of BH in
Galactic binaries and those in AGNs.

CG~X-1 was first observed by $ROSAT$/HRI in 1995 \citep[source 3
in][]{Guainazzi99}, with a 0.5-2 keV flux of $\simeq2\times10^{-13}$ erg
cm$^{-2}$ s$^{-1}$, fully consistent with that of the low state of the June 2000
$Chandra$ observation. The Circinus Galaxy was observed again by $ROSAT$/HRI in the
following years: we downloaded the data from the archive and analysed the count
rates of CG~X-1 for each of them (Table \ref{rosat}). There is evidence of some
variability, but the source seems always to be caught in a low flux state.
On the other hand, a short-term, periodic variability is observed in
the observations with the highest exposure times. The clearest evidence comes
from the longest observation, in August 1997, whose lightcurve power spectrum
shows a peak at a frequency corresponding to a period of $\simeq 27$ ks, as
observed by \citetalias{Bauer01} in $Chandra$ data (see below). The 1990
$ROSAT$/PSPC observation could not be used to study CG~X-1 because of its poorer
spatial resolution ($\simeq25\arcsec$), which is actually larger than the
distance between the source and the nucleus.

\begin{table}

\caption{\label{rosat}Count rates (0.5-2 keV) of CG~X-1 for all $ROSAT$/HRI
observations. The 0.1-10 keV fluxes are inferred from the powerlaw model
parameters of the $Chandra$ June 2000 observation.}

\begin{center}

\begin{tabular}{llll}
\hline \textbf{Date} & \textbf{Exp. (s)} & \textbf{c/s} & \textbf{Flux 0.1-10 keV
(cgs)}\\
\hline
SEP95 & 4193 & $0.0045\pm0.0010$ & $9.5\times10^{-13}$\\
FEB96 & 1081 & $0.0019\pm0.0013$ & $4.0\times10^{-13}$\\
SEP96 & 1850 & $0.0011\pm0.0008$ & $2.3\times10^{-13}$\\
MAR97 & 26673 & $0.0023\pm0.0003$ & $4.8\times10^{-13}$\\
AUG97 & 46367 & $0.0027\pm0.0002$ & $5.7\times10^{-13}$\\
\hline
\end{tabular}

\end{center}

\end{table}

CG~X-1 was then observed by $Chandra$ and studied in detail by \citet{sw01},
\citet{Sambruna01a} and, in particular, \citetalias{Bauer01}. Indeed,
the best spectral information come from the longest observation, in June 2000,
which also gives the clearest evidence of a short-term, periodical behaviour of
the source \citepalias{Bauer01}. Therefore, we will mainly refer to the results
presented by \citetalias{Bauer01} and to their proposed
interpretations of the nature of CG~X-1, adding further insights using
our analysis of all $Chandra$ observations (see Table \ref{Jfit}). Moreover,
since it was shown in Sect. \ref{spectral} that this source is likely to be
the origin of most of the flux excess in the second BeppoSAX observation (with
the exception of the 6.7 keV line and the PDS variation: see next
section), we can assume that the spectrum resulting from the subtraction of the
two BeppoSAX observations is a genuine spectrum of CG~X-1. This spectrum can
give valuable information, because, together with the $Chandra$ June 2000
observation, is the only one which includes more than a period of the source.

There are several pieces of evidence which strongly suggest that CG~X-1 lies
in the disk of the Circinus Galaxy \citepalias{Bauer01}. At the distance of 3.8 Mpc,
the flux peak measured in March 2000 corresponds to a luminosity of
$\sim10^{40}$ erg s$^{-1}$: this implies an object with a mass M $\geq$ 80
M$_{\sun}$, if radiating at or below its Eddington limit.

An optical counterpart is not detectable down to $m_\mathrm{V} < 25.3$ mag
\citepalias{Bauer01}. One of the most characteristic feature of CG~X-1 is its
clear periodic behaviour: in the June 2000 observation it was found to display a
flux variability by a factor $\simeq20$ in a period of $7.5\pm0.2$ hr
\citepalias{Bauer01}. However, long-term variability is also observed, as
already pointed out in Sect. \ref{chandraflux}: the flux peak in the June
observation, for example, is about 1/4 of the observed mean in March. Periodic
behaviour was also found in our analysis of the high flux state of March 2000,
but the shorter exposure time prevented from confirming the above-mentioned
period. The shape of the lightcurve during a period could be simply explained
with an eclipse by a binary companion, but in order to make this model
self-consistent an unphysically large companion would be required
\citepalias{Bauer01}.

Another possibility, again suggested by \citetalias{Bauer01}, is the modulation
of the accretion rate from accretion-disk instabilities, as observed in the
quasi-periodic episodes of the Galactic microquasar GRS~1915+105
\citep{gmr96,Belloni97}. The statistical quality of the available
observations of CG~X-1 is not good enough to perform a spectral variability
analysis as detailed as the one performed by
\citet{Belloni97}, which led them to explain the variability in GRS~1915+105 in
terms of cyclical emptying and refilling of the inner accretion disk. However,
\citetalias{Bauer01} claim a possible variation of the spectral shape during the
different phases of the periodical lightcurve and the different temperatures
needed in our fits of all $Chandra$ observations of CG~X-1 (see Table \ref{Jfit})
could be due to the fact that we are observing the source in different
phases of the period.

Regarding the long-term variability of CG~X-1, if we compare the only
two observations with a spectral shape averaged over more than a period
(BeppoSAX and $Chandra$ June 2000), we see that, despite a change in flux of a
factor $\simeq6$, the temperatures for the disk black body model are consistent
with each other (see Table \ref{Jfit}). This suggests that the long term
variability is not associated with significant variations of the
inner radius of the accretion disc.

\subsection{The ionized iron line and the PDS variation}

The spectral and flux properties of CG~X-1 are clearly the best
explanation for the variations observed in the Circinus Galaxy. However, no iron
line is present in the $Chandra$ spectrum of CG~X-1, while large residuals are
clearly detected in the new BeppoSAX data in excess of the iron line observed in
the old data. The centroid energy of the residuals is strongly shifted to 6.7
keV, indicating ionized iron. The flux required by the SAX data is much larger
than the upper limit in the spectrum of CG~X-1, also when the source was as
bright as supposedly was in BeppoSAX.

We re-analysed data from the old BeppoSAX observation and added a line at
6.67 keV: the upper limit for its flux is  $2.6\times10^{-5}$ ph
cm$^{-2}$s$^{-1}$, which is indeed a factor of 3 lower than in the new data
(Sect. \ref{saxspectral}). On the other hand, the above-mentioned upper limit is 
fully consistent with the flux observed by the gratings aboard $Chandra$
\citep[$2.7\times10^{-5}$ ph cm$^{-2}$s$^{-1}$:][]{Sambruna01b} in the nuclear
region.

Contamination from another off-nuclear source is a possible explanation.
CG~X-2 \citepalias[see Fig. 1 in ][]{Bauer01} is an object with a flux 
comparable to that of CG~X-1 in its low state, but fairly constant during all
$Chandra$ observations, being $F_{0.1-10 \rm \, keV}=1.3\times10^{-12}$ erg
cm$^{-2}$ s$^{-1}$. Since it was completely absent in the 1995 $ROSAT$/HRI
observation and has optical and radio counterparts, \citetalias{Bauer01}
suggested it could be a young SNR. Our analysis of the other $ROSAT$/HRI
observations shows the source is not present up to at least September 1997, i.e.
six months before the first BeppoSAX observation.

As observed by \citetalias{Bauer01}, CG~X-2 has a huge iron line (EW=1.59 keV) at
$\approx6.8$ keV which is likely the blend of two lines (at 6.7 and 6.97 keV),
as suggested by its intrinsic width of 190 eV. The flux of the line, which we
calculated in the longest $Chandra$ observation, is $1.6\pm{0.4}\times10^{-5}$
ph cm$^{-2}$ s$^{-1}$, which is almost a factor 5 lower than required in
the new BeppoSAX data. The shorter exposure times of the other observations do
not allow to search for variability of the flux of the line.

Therefore, it is not possible to exclude that the ionized iron line observed in
the second SAX observation, even if clearly including the line of CG~X-2 that
was probably not present in the previous observation, could be an effect of a
variation in the nuclear emission.

The same explanation can be adopted for the variation in the PDS spectral shape.
Although the total count rates are consistent with each other (see Table
\ref{saxcount}), there is clearly a decrease in the lower part of the PDS band
(Fig. \ref{new2old}). A possible interpretation is in terms of a change of the
column density of the absorbing matter (the torus). However, it must be noted
that the PDS field of view is very large (1.5x1.5 degrees) and so the emission
from the nucleus could be contaminated by a number of sources in an
unpredictable way. Among these sources, an important contribution could come
from the second source present in the LECS and MECS images (see
Sect. \ref{saxflux}): it may be that its flux over 10 keV was higher in the
first SAX observation. Unfortunately, there is no way to test this hypothesis.

\section{Conclusions}

The most likely explanation of at least most of the flux and spectral variation
observed between two BeppoSAX observations taken almost three years apart is in
terms of a high-flux state of CG~X-1, which is located well within the adopted
BeppoSAX source extraction region of $2\arcmin$. There are several pieces of
evidence:

\begin{itemize}
\item CG~X-1 shows a strong, long-term variation, reaching a flux which is
comparable with that of the nucleus and is consistent with that measured in the
residuals of the two SAX observations.
\item The spectral shape of CG~X-1, when averaged over more than a period, is
fully consistent with the spectrum of the LECS and MECS BeppoSAX residuals.
\item Finally, the strongest evidence comes from the short-term variation of
CG~X-1, on a period of 27 ks: the MECS lightcurve of the new observation
clearly varies with the same period, indicating that a significant part of the
observed flux originates from CG~X-1. The periodic behaviour was also present
in the data from the old SAX observation, but the variation amplitude was much
less, being consistent with contamination from a low-flux state of CG~X-1.

\end{itemize}

The best interpretation for the nature of CG~X-1 is in terms of a $\geq$ 80
M$_{\sun}$ black hole in an accreting binary system in Circinus, as previously
suggested by \citetalias{Bauer01}. Our analysis of all $Chandra$
observations and the BeppoSAX spectrum shows that the long-term variability
of this source is not associated to spectral variations, indicating
that it is not due to changes of the inner radius of the disk.

However, CG~X-1 cannot be the cause of the variation of the PDS and the
presence of an ionized iron line in the residuals between the two BeppoSAX
observations. In both cases, a variation of the properties of the circumnuclear
matter in the AGN environment is a possible explanation: an increase of the
column density of the torus, for example, would cause the observed decrease of
the PDS count rates at lower energies. However, it is difficult to imagine the
physical cause to support such an increase.

As for the ionized iron line, it should be noted that the SAX residuals must
include a line from CG~X-2, which was likely not present in the old
observation. Nevertheless, the contribution of the flux of this line is too low
to exclude a variation of the 6.7 keV iron line originating from the nucleus.

Finally, our new analysis of the $Chandra$ HEG nuclear spectrum has led to the
detection of a Compton shoulder in the 6.4 keV iron line. Its flux is about
20$\%$ the line core, in agreement with theoretical expectations for
Compton-thick matter, providing one more argument in favour of the
association of the matter producing the iron line with the $4\times10^{24}$
cm$^{-2}$ neutral absorber.

\begin{acknowledgements}
We would like to thank Harvey Tananbaum for his generous grant of a Chandra
DDT observation, and the anonymous referee for his valuable suggestions which
helped us improving the clarity of the paper.
\end{acknowledgements}

\bibliographystyle{aa}
\bibliography{sbs}

\end{document}